\documentclass[preprint,showpacs,preprintnumbers,amsmath,amssymb,superscriptaddress,bibnotes, aps, pra]{revtex4-1}

\usepackage{graphicx}% Include figure files
\usepackage{dcolumn}% Align table columns on decimal point
\usepackage{tabularx}
\usepackage{bm}% bold math
\usepackage{color}
\makeatletter
\def\Ddots{\mathinner{\mkern1mu\raise\p@	
		\vbox{\kern7\p@\hbox{.}}\mkern2mu
		\raise4\p@\hbox{.}\mkern2mu\raise7\p@\hbox{.}\mkern1mu}}
\makeatother

\begin{document}
	
	\preprint{Preprint}
	\title{Interaction of carrier envelope phase-stable laser pulses with graphene: the transition from the weak-field to the strong-field regime}
	
	\author{Christian Heide} 
	\email[E-mail: ]{christian.heide@fau.de}
	\affiliation{Laser Physics, Department of Physics, Friedrich-Alexander-Universit\"at Erlangen-N\"urnberg, Staudtstrasse 1, D-91058 Erlangen, Germany}
	\author{Tobias Boolakee} 
	\affiliation{Laser Physics, Department of Physics, Friedrich-Alexander-Universit\"at Erlangen-N\"urnberg, Staudtstrasse 1, D-91058 Erlangen, Germany}
	\author{Takuya Higuchi}
	\affiliation{Laser Physics, Department of Physics, Friedrich-Alexander-Universit\"at Erlangen-N\"urnberg, Staudtstrasse 1, D-91058 Erlangen, Germany}
	\author{Heiko B. Weber}
	\affiliation{Applied Physics, Department of Physics, Friedrich-Alexander-Universit\"at Erlangen-N\"urnberg, Staudtstrasse 7, D-91058 Erlangen, Germany}
	\author{Peter Hommelhoff}
	\email[E-mail: ]{peter.hommelhoff@fau.de}
	\affiliation{Laser Physics, Department of Physics, Friedrich-Alexander-Universit\"at Erlangen-N\"urnberg, Staudtstrasse 1, D-91058 Erlangen, Germany}
	\date{\today}% It is always \today, today,
	
\begin{abstract}
Ultrafast control of electron dynamics in solid state systems has recently found particular attention. By increasing the electric field strength of laser pulses, the light-matter interaction in solids might turn from a perturbative into a novel non-perturbative regime, where interband transitions from the valence to the conduction band become strongly affected by intraband motion. We have demonstrated experimentally and numerically that this combined dynamics can be controlled in graphene with the electric field waveform of phase-stabilized few-cycle laser pulses \cite{Higuchi2017, Heide2018}. Here we show new experimental data and matching simulation results at comparably low optical fields, which allows us to focus on the highly interesting transition regime where the light-matter interaction turns from perturbative to non-perturbative. We find a 5th order power-law scaling of the laser induced waveform-dependent current at low optical fields, which breaks down for higher optical fields, indicating the transition.
\end{abstract}%
\maketitle
\tableofcontents
\section{Introduction}
\begin{figure}[t]
	\begin{center}
		\includegraphics[width=16cm]{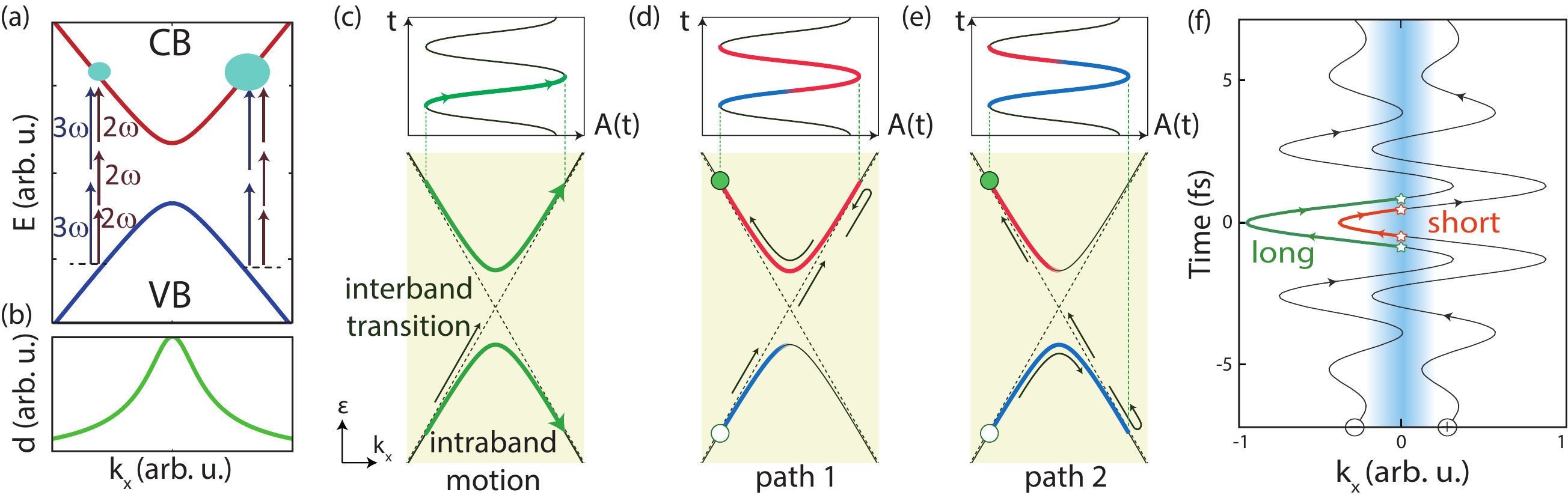}% Here is how to import EPS art
		\caption{\label{Figure1} (a) Illustration of the band structure of graphene near the $K$ point with the conduction band (CB) in red and the valence band (VB) in blue. Quantum-path interference between two-photon absorption and three-photon absorption near the K point of graphene is depicted. When the two pulses are phase-coherent, the quantum-path interference can be constructive, resulting in an excitation of the electron from the VB to CB or destructive, resulting in no excitation. (b) The interband dipole moment $d$ is maximized near the band gap minimum $k_\text{x} =0$ (schematic illustration). (c) Combined intraband motion and interband transition of an electron in the reciprocal space, driven with the optical field. When an electron approaches the bandgap minimum via intraband motion, a Landau-Zener transition (LZ) can occur with a certain probability $P_\text{LZ}$, driving the electron from one band to the other. Within one optical cycle two LZ transition can occur. (d),(e) Two quantum pathways with the same initial and final state. The electron can first undergo an interband transition followed by intraband motion (d), or vice versa (e). (f) Electron trajectory in the reciprocal space driven by the light field. The blue-shaded line represents the region with the highest interband transition probability, with the white stars indicating the main transition events. For an electron starting with positive (negative) wavenumber, the electron trajectory after passing the LZ transition is short (long), which results in a different accumulated dynamical phase and subsequently a different interference condition (see \cite{Higuchi2017}).} 
	\end{center}
\end{figure}

Interactions between intense optical fields and matter have facilitated controlling electron trajectories coherently on attosecond timescales \cite{Ghimire2011, Kruger2012, Schultze2013, Hohenleutner2015, Vampa2015a, Vampa2015, Luu2015, Garg2016, Higuchi2017, Yoshikawa2017, Liu2017, Garzon-Ramirez2018, Reimann2018, Langer2018, Heide2018}. In particular, the electron dynamics can be controlled by the exact shape of the waveform of ultrashort laser pulses, enabling sub-optical-cycle manipulation such as in high-harmonic generation in gases and solids \cite{Ghimire2011, Hohenleutner2015, Vampa2015a, Vampa2015, Luu2015, Liu2017, Yoshikawa2017}, light-field-driven ionization and electron emission \cite{Kruger2012}, and light-field-driven current generation in solids \cite{Schultze2013, Garg2016, Higuchi2017, Garzon-Ramirez2018, Reimann2018, Heide2018}.\\
In the limit of weak optical fields (i.e. $E\ll1$~V/nm for laser pulses with a centre wavelength of about 1~$\mu$m and materials with a eV-level work function or band gap), the change in the electron wavenumber during the light-matter interaction can be neglected and the interaction can be treated perturbatively. In this regime perturbative photon absorption is appropriate to describe the light-matter interaction. It has been shown that multi-photon quantum-path interference between different excitation pathways can generate an electrical current in solids \cite{Atanasov1996, Hache1997,Laman1999,Fortier2004, Sun2010, Muniz2018, Wang2018}, control the electron emission from metal surfaces \cite{Forster2016} or the ionization process in atoms \cite{Muller1990}. This quantum-path interference is sensitive to the optical phase between different photon energies, which can be controlled with the electric field waveform of ultrashort laser pulses or by the temporal delay between two-colour laser fields. Figure~\ref{Figure1}(a) illustrates exemplary odd-order perturbation at an avoided crossing, such as 2$\omega$-3$\omega$ interference, which can result in a asymmetric conduction band population \cite{Wang2018}. Here, $\omega$ denotes the mean carrier frequency, 2$\omega$ the second harmonic and so on.\\
In contrast, under an intense electric field strength (i.e. $E\gg1$~V/nm) the change in the electron wavenumber during the light-matter interaction can no longer be neglected. Due to the change of the electrons' wavenumber the electrons may pass nearby an avoided crossing formed around band gaps in solids \cite{Vampa2015a, Chizhova2016, Higuchi2017, Heide2018, Schlaepfer2018, Sato2018}. When the electron approaches nearby the avoided crossing the dipole moment between the two coupled bands is maximal, resulting in an enhanced transition probability from one band to the other. This process is known as a Landau-Zener transition \cite{Zener1932, Shevchenko2010} (see Fig.~\ref{Figure1}(b), Fig.~\ref{Figure1}(c)).
\section{Recapitulation of strong-field physics in graphene}
We have recently reported that the interplay of intraband motion and interband transition results in a residual electric current, which can be controlled with the electric field waveform of few-cycle laser pulses, on attosecond timescales \cite{Higuchi2017, Heide2018}. The waveform of the optical pulses is characterized by the carrier-envelope phase (CEP), which determines where the maximum of the carrier field lies within the pulse envelope. We have found a CEP-dependent current in a graphene stripe, excited with linearly polarized laser pulses with a peak optical field of up to 3~V/nm. The amplitude of the current scales strongly nonlinearly with field strength and shows a current reversal at around 2~V/nm.\\
We interpret this change in direction of the current as a result of a transition from the perturbative (photon-driven) to the non-perturbative (optical-field-driven) regime.\\
In the optical-field-driven regime, the wavenumber \textbf{k}(t) of the electrons is changed by the optical field
\begin{align}
\label{eq. acceleration theorem}
\frac{\text{d}\textbf{k}(t)}{\text{d}t} = \frac{e\textbf{E}(t)}{\hbar}
\end{align}
resulting in intraband motion. When the electric field $\textbf{E}(t) = \textbf{E}_\text{0}$ is static, $\textbf{k}(t)$ changes linearly as a function of time. In this case, the transition probability for electrons to go from the valence band to the conduction band can be well described by the Landau-Zener (LZ) framework \cite{Zener1932, Shevchenko2010}. Under an oscillating electric field, the electrons in graphene may undergo repeated LZ transitions. As shown in Fig.~\ref{Figure1}(c), the electron undergo either an interband LZ transition from one band to the other, or it performs an intraband motion when passing $k_\text{x}$=0. Within one optical cycle of the electric field waveform two transition events can occur: First the electron might undergo an LZ transition and then intraband motion (Fig.~\ref{Figure1}(d)) or vice versa (Fig.~\ref{Figure1}(e)). Since this process is faster than any dephasing in graphene, these two indistinguishable quantum pathways with the same initial and final state can interfere. This process, comprised of two subsequent coherent Landau-Zener transitions, is known as Landau-Zener-St\"uckelberg interferometry \cite{Shevchenko2010}.\\
Although graphene's band structure is inversion symmetric around $k_\text{x}$=0, the electric field waveform of few-cycle laser pulses can break this symmetry. Figure~\ref{Figure1}(f) shows the corresponding electron trajectory in the reciprocal space for an electron with positive and negative initial wavenumber. Depending on the initial wavenumber of the electron the phase accumulation during the electron trajectory might be different.\\
Numerical simulation results suggest that the excitation probability from the valence band to the conduction band is well described as a result of interference of different quantum-pathways with different phase accumulations. This interference is sensitive to the temporal evolution of the electric field, which explains quantitatively the experimentally observed carrier-envelope-phase dependence of the photocurrent under few-cycle laser pulse illumination \cite{Higuchi2017}.\\
Since the change in the electrons' wavenumber is proportional to the electric field strength, the length of the electrons' trajectory in the reciprocal space and thus the phase accumulation can be controlled with the magnitude of the electric field strength. Increasing the length of the electron trajectories might change the interference condition form constructive to destructive or vice versa. Our simulation suggests that the first change in the interference condition is found at around 2~V/nm (see \cite{Higuchi2017} for graphene and also \cite{Wismer2016} for GaAs).
\section{Experimental methods and results}
Here we show new data with higher quality and more data points at low optical fields ($E<$ 2~V/nm), which allows us to have a deeper look into the transition regime of perturbative and non-perturbative light-matter interaction. Furthermore we will show the full measured data set in insightful polar diagrams and compare them with numerical simulations. The non-perturbative regime for $E>$2~V/nm has been discussed in detail in \cite{Higuchi2017}, so we will focus on the perturbative regime here. In the following we will briefly outline the main experimental setup and techniques. A more detailed discussion can also be found in \cite{Higuchi2017, Heide2018}.\\
We perform photocurrent measurements in graphene by illuminating it with CEP-controlled laser pulses for different electric field strengths for linearly and circularly polarized pulses. We use monolayer graphene epitaxially grown on 4H silicon carbide (SiC) \cite{Emtsev2009}. A graphene stripe with a width of 2$~\mu$m and a length of 5$~\mu$m is patterned by electron-beam lithography and plasma etching. Two gold electrodes with titanium as adhesive layers are deposited to detect the electric current. We focus near-infrared laser pulses with a Fourier-transform limited pulse duration of about 5.4~fs (FWHM of intensity envelope) with an off-axis parabolic mirror to a focus with a beam waist of $w_0 = 1.5~\mu$m to the centre of the graphene stripe. To pick up the current contribution, which is sensitive to the waveform of the laser pulses, we modulate the carrier-envelope phase (CEP) periodically by setting the carrier-envelope-offset frequency to $f_\text{CEO} = \frac{\text{d}}{\text{d}t}$(CEP) = 1.1~kHz. The current amplitude $J$ and the phase $\Phi_\text{J}$ with respect to $f_\text{CEO}$ is detected using a lock-in detection scheme. For each field strengths and polarization state, the so-obtained CEP-dependent current was measured for 20~s with a integration time of 300~ms to obtain the mean values. Higher harmonics of the modulation frequency are not considered here. 

\begin{figure}[t]
	\begin{center}
		\includegraphics[width=12cm]{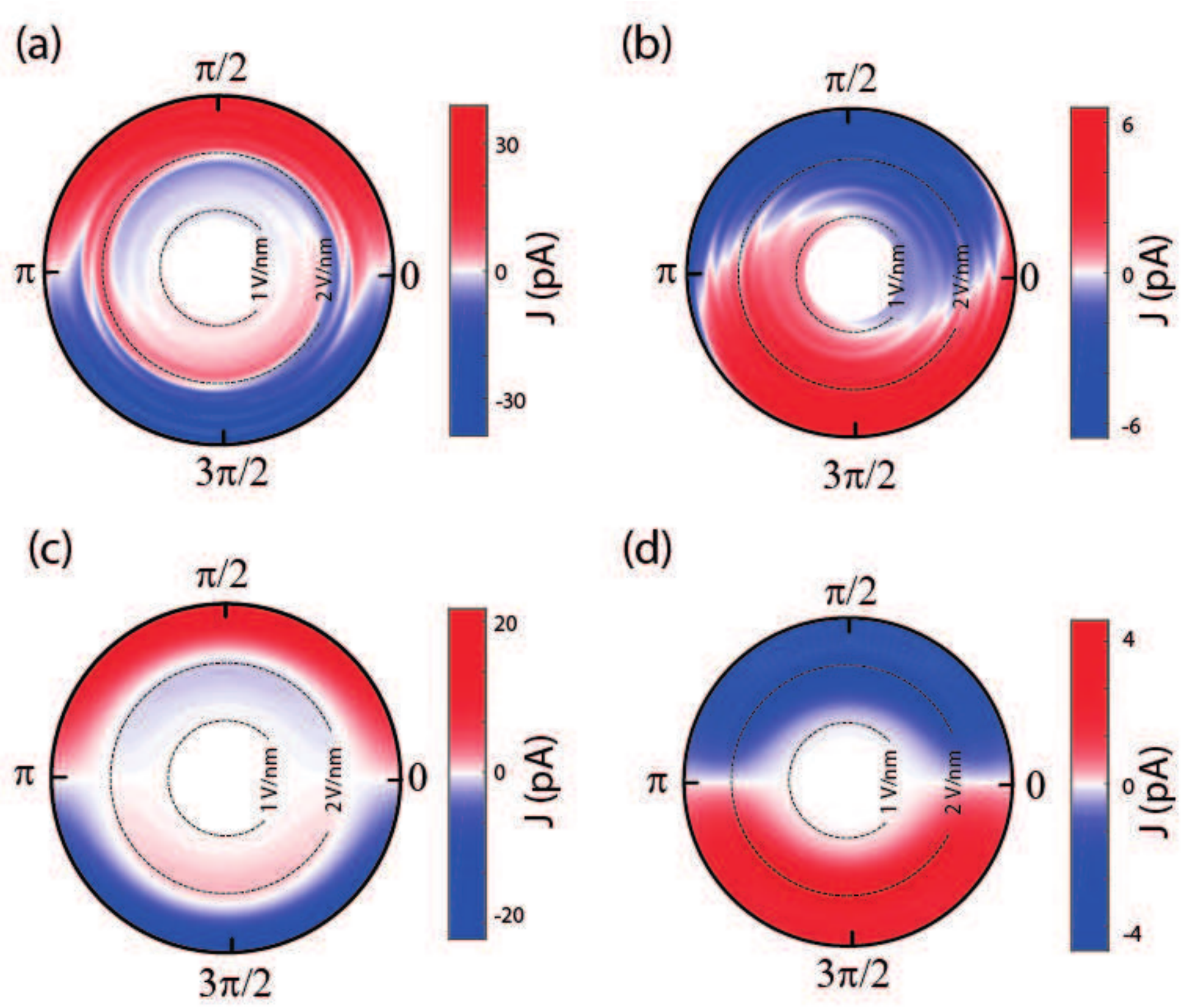}% Here is how to import EPS art
		\caption{\label{Figure4} (a) Measured CEP-dependent current using linearly polarized light. In these polar diagrams, the radius corresponds to the peak electric field strength $E$, which varies from 0.5~V/nm to 3~V/nm, while the polar angle encodes the carrier-envelope phase. The colour coding represents the CEP-dependent current, measured using a lock-in technique. (b) Polar plot for circularly polarized light. (c),(d) Numerical simulation results of the residual current for linearly and circularly polarized light, respectively. The main features of the experimental data are all quantitatively reproduced. See text for details.}
	\end{center}
\end{figure}

The polar diagrams in Figs.~\ref{Figure4}(a) and \ref{Figure4}(b) show the complete measurement set of the CEP-dependent current for linearly and circularly polarized light. The radius indicates the peak electric field strength $E$, which varies from 0.2~V/nm to 3~V/nm. Since the graphene is on the surface of SiC, the optical field strength applied to graphene rests reduced compared to the bare focus in vacuum by a factor $\frac{2}{1+n_\text{SiC}}$, with $n_\text{SiC}$ the refractive index of SiC at the laser centre wavelength. The electric field strength $E$ we give includes this factor. The polar angle in Figs.~\ref{Figure4}(a)-\ref{Figure4}(b) encodes the measured phase $\Phi_\text{J}$. Based on our numerical simulations discussed in \cite{Higuchi2017} we have calibrated this phase by maximizing the CEP-dependent current for a CEP of $\pi/2$ at a field strength of 3~V/nm. We keep this phase calibration constant for all $E$ and polarization states. The color coding represents the measured CEP-dependent current.

We see in Fig.~\ref{Figure4}(a) that for an electric field strength below 2~V/nm the current generated by laser pulses with $\Phi_\text{CEP} = \pi/2$ is negative and its magnitude increases monotonically as a function of $E$. Around a field strength of 2~V/nm a sudden $\pi$ jump in $\Phi_\text{J}$ reverse the CEP-dependent-current direction, resulting in a positive current for $\Phi_\text{CEP}= \pi/2$. For $\Phi_\text{CEP}$ = 0 or $\pi$ almost no CEP-dependent current can be found ($<$ 5$\%$ of $J(\pi/2)$). For the case of circularly polarized light the current increases monotonically and no change in current direction can be found. Note that rotating the quarter waveplate by $45^{\circ}$ to obtain circular polarization from linear polarization introduces a phase shift of $\pi/4$ in the CEP. The presented data are corrected for this phase shift.\\
Figures~\ref{Figure4}(c) and \ref{Figure4}(d) show the simulation data introduced in \cite{Higuchi2017}, but here in polar diagrams. Clearly, the main features of the experimental results match quite well. In particular, the change in current direction for linear polarisation and its absence for circular polarization are fully reproduced, also quantitatively. Deviations between simulation and experiment could originate from the simplification of the laser spectrum as a Gaussian spectrum or the assumption of a pure two-level system in graphene. Both of these simplifications are necessary given our limited computing resources. More in-depths simulation work is needed to obtain a more detailed understanding. Interestingly, Wismer \textit{et al.} simulated how the number of bands considered as well as the dephasing during the light-matter interaction in GaAs might influence the CEP-dependent current generation process \cite{Wismer2016}. Their results show how multi-band effects change the shape of the polar diagrams, with results that seem to indicate similar effects to arise for our simulation data. Because of the different band structure of the investigated systems, this hint needs to be taken with care.

\begin{figure}[t]
	\begin{center}
		\includegraphics[width=12cm]{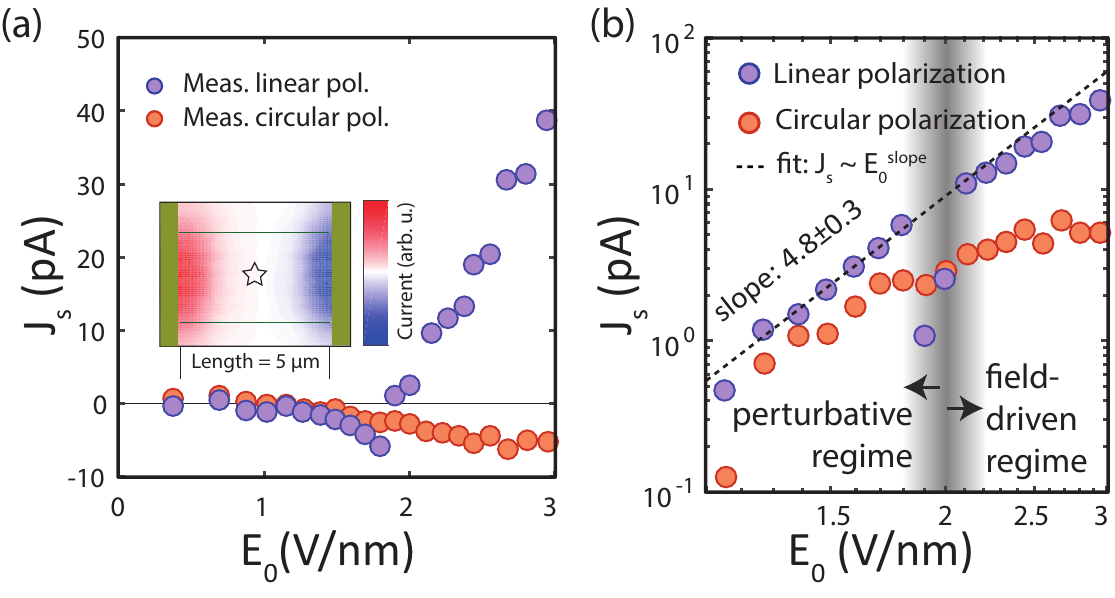}% Here is how to import EPS art
		\caption{\label{Figure5} 
			CEP-dependent residual current (a), Difference in current generation for $\Phi_\text{CEP} = - \pi/2$ and $\Phi_\text{CEP} = \pi/2$ are plotted for linearly and circularly polarized light excitation. For each data point, the signal was integrated over 20~s to obtain the mean values plotted. The standard error of the mean over the integration time is smaller than 8 percent of the mean value for all data points. For circular polarization, the peak field strength along both, the $x-$ and $y-$ axis are $E/\sqrt2$. The inset shows the photocurrent map of the graphene stripe. The electrodes are colored in gold. For the measurement the position, marked with a star is illuminated. (b) Current as a function of the field strength plotted double logarithmic. In the weak-field regime, the power law scaling obtained from the fit is 4.8$\pm$0.3.}
	\end{center}
\end{figure}
Figure~\ref{Figure5}(a) shows the sinusoidal part of the CEP-dependent current
\begin{align}
J_s(E) = J\sin\left(\Phi_\text{j}\right)
\end{align}
for linearly and circularly polarized illumination as a function of the peak electric field strength $E$. Hence, $J_s$ corresponds to the ($\pi/2$)-($3\pi/2$)-axis in Fig.~\ref{Figure4}(a) and \ref{Figure4}(b). Increasing the electric field strength for linearly polarized light from 0.2~V/nm to 1.8~V/nm results first in a negative $J_s$ with a superlinear increase as a function of $E$, which suddenly turns around 2~V/nm into a positive $J_s$, where it stays up to the maximally attainable field strength of 3~V/nm. For circularly polarized light $J_s$ increases first superlinearly and starts to saturate at around 2~V/nm. Importantly, it stays negative up to 3~V/nm.\\
To gain insight into the superlinear regime at weak optical fields, we plot $J_s$ versus $E$ in double logarithmic plots in Fig.~\ref{Figure5}(b). We can see that below $E=$ 1.8~V/nm the current follows a $E^5$-scaling, with deviations above 2~V/nm.

As discussed in the introduction, it has been shown that also multi-photon quantum-path interference can generate a CEP-dependent excitation in the perturbative regime \cite{Hache1997, Sun2010}. Quantum-path interference of odd-order perturbation such as $\omega- 2\omega$ or $2\omega - 3\omega$ can result in an asymmetric electron population and hence a current generation. This current rises with increasing laser power. In our measurement the employed laser spectrum spans from 600~nm to 1150~nm and does not contain $\omega$ and 2$\omega$ frequency components within its bandwidth (measured down to -40 dBc). Therefore, an interference between one-photon absorption of a 2$\hbar\omega$-photon and two-photon absorption of $\hbar\omega$ is suppressed, but higher order processes like $2\omega$ - $3\omega$ may still lead to quantum-path interference. For $2\omega$ - $3\omega$ quantum-path interference an $E^5$-scaling is expected \cite{Wang2018}. Our power scaling data in the perturbative regime can thus be well explained with 2-vs-3-photon quantum-path interference, showing the tell-tale $E^5$ behavior of $J_s$ (Fig.~\ref{Figure1}\textbf{a}).\\
In the strong-field regime such a perturbation picture is no longer applicable \cite{Wachter2015, Wismer2016}. In these works, it has been shown that a current reversal together with the deviation in the power-law scaling of the current as a function of the electric field strength may well be interpreted as a transition from the weak-field (perturbative light-matter interaction) to the strong-field (non-perturbative light-matter interaction) regime.
\section{Conclusion and outlook}
In summary, we have presented experimental data and simulation results clearly showing a transition from the perturbative to the non-perturbative regime at a field strength of 1.8~V/nm in graphene driven with 2-cyle laser pulses at a centre wavelength of 800~nm.
For higher optical fields intraband motion strongly affects interband transitions, and a perturbative picture (photon absorption) is no longer appropriate to describe the light-matter interaction. We have identified the transition between the two regimes via a deviation from 5th order power-law scaling at small field strengths, evidencing a five-order multi-photon process. In addition, we find a peculiar change in the current direction when crossing the transition field strength. In the strong-field regime, we find field-driven Landau-Zener-Stückelberg interference, as discussed in detail previously. This transition at an electric field strength of 1.8~V/nm is characteristic for graphene's band structure and the applied laser parameter. Since LZS interference is not limited to graphene's band structure, it might also be found in the electron dynamics around band gap in various different solid state systems \cite{Vampa2015a, Higuchi2017}. There the transition region might be found at different characteristic field strengths.
\section*{Acknowledgments}
This work has been supported in part by the European Research Council (Consolidator Grant NearFieldAtto) and Deutsche Forschungsgemeinschaft (DFG, German Research Foundation) Projektnummer 182849149 – SFB 953. 
\providecommand{\newblock}{}

\end{document}